\documentclass[conference,10pt]{IEEEtran}
\usepackage[T1]{fontenc}
\usepackage[utf8]{inputenc}
\usepackage{amssymb}
\usepackage{amsmath}
\usepackage{array}
\usepackage{latexsym}
\usepackage{cite}

\usepackage{amsmath,amsfonts,bm}

\def\eqref#1{equation~\ref{#1}}

\def\1{\bm{1}}

\DeclareMathAlphabet{\mathsfit}{\encodingdefault}{\sfdefault}{m}{sl}
\SetMathAlphabet{\mathsfit}{bold}{\encodingdefault}{\sfdefault}{bx}{n}

\usepackage{float}
\usepackage{graphicx}
\usepackage{wrapfig}
\usepackage{booktabs}
\usepackage{stfloats}
\usepackage{subcaption}

\usepackage[table]{xcolor}
\usepackage{caption}
\definecolor{rowblue}{RGB}{231,243,255}

\usepackage[shortlabels]{enumitem}

\usepackage{url}
\usepackage{comment}

\usepackage{microtype}
\usepackage{inconsolata}

\usepackage{xcolor}

\newcommand\be[1]{\textbf{\emph{#1}}}

\newcommand\anurag[1]{{\color{red}{[Anurag: #1]}}}

\newcommand{\eat}[1]{}

\title{Scaling State-Space Models on Multiple GPUs with Tensor Parallelism}

\author{
\IEEEauthorblockN{Anurag Dutt}
\IEEEauthorblockA{
Stony Brook University\\
adutt@cs.stonybrook.edu
}
\and
\IEEEauthorblockN{Nimit Shah}
\IEEEauthorblockA{
Stony Brook University\\
nimishah@cs.stonybrook.edu
}
\and
\IEEEauthorblockN{Hazem Masarani}
\IEEEauthorblockA{
Stony Brook University\\
hazem.masarani@stonybrook.edu
}
\and
\IEEEauthorblockN{Anshul Gandhi}
\IEEEauthorblockA{
Stony Brook University\\
anshul@cs.stonybrook.edu
}
}

\begin{document}

\maketitle

\begin{abstract}


Selective state space models (SSMs) have rapidly become a compelling backbone for large language models, especially for long-context workloads. Yet in deployment, their inference performance is often bounded by the memory capacity, bandwidth, and latency limits of a single GPU, making multi-GPU execution increasingly necessary. Although tensor parallelism (TP) is widely used to scale Transformer inference, applying it to selective SSM blocks is non-trivial because the SSM mixer couples large projections with a sequence-wise recurrent state update and local mixing whose efficiency depends on preserving locality and avoiding synchronization in the critical path.

This paper presents a communication-efficient TP design for selective SSM inference that addresses three practical engineering challenges: enabling TTFT improvements via an SSM state cache across prefill and decode, partitioning the mixer’s packed parameter tensor so that recurrent updates remain local while minimizing communication, and reducing TP aggregation overhead with quantized all-reduce. We evaluate on three representative SSM-based LLMs spanning pure-SSM and hybrid architectures---Mamba, Falcon-Mamba, and Zamba---on NVIDIA A6000 and A100 clusters. Our experiments show substantial throughput gains from tensor-parallel SSM inference, improving batch-request throughput by $\sim$1.6--2.1$\times$ on 2 GPUs and $\sim$2.6--4.0$\times$ on 4 GPUs for Mamba, with the largest benefits at long context lengths, and achieving a further $\sim$10--18\% throughput improvement from quantized all-reduce by lowering synchronization bandwidth overhead.
\end{abstract}



\section{Introduction}
\label{s:introduction}
\textbf{State Space Models (SSMs)} have recently emerged as a strong backbone for sequence modeling, with competitive results and rapid adoption in large language modeling~\cite{gu2023mamba, jamba2024, visualMambaSurvey, localmamba}. 
A key appeal is their ability to handle long contexts efficiently: instead of forming pairwise token interactions as in the attention mechanism in Transformer models, SSM layers process tokens in order while maintaining a compact internal \emph{state} that summarizes prior context~\cite{gu2023mamba,dao2025transformersareSSMs}.
SSM mixers also scale \emph{linearly} in sequence-processing costs compared to transformers (self-attention), which are \emph{quadratic}.
An SSM block processes input tokens by continuously updating the internal state as it reads tokens, and producing an output for each token from this evolving state memory; the way the memory is updated depends on the current token, which makes SSMs flexible and improves modeling quality at scale.
We discuss SSMs in detail in Section~\ref{ss:bg_ssm_arch}.


As SSMs move to deployment-relevant model sizes, the systems focus shifts to \emph{improving serving performance}: maximizing throughput under the memory and bandwidth constraints of the deployment. 
In particular, as SSM models grow and serving stacks support longer contexts and higher concurrency, deployments must simultaneously (i) fit model weights and runtime buffers, and (ii) sustain high token throughput.
However, despite their advantages and flexibility, the serving performance of SSMs in practice is largely constrained by the limits of a single GPU (memory capacity, compute capabilities).
This is because of the \be{lack of an established GPU-parallelized implementation} for SSMs.
As model sizes continue to grow, the lack of a multi-GPU SSM implementation becomes increasingly limiting.

The common approach to harnessing the resources of multiple GPUs is through parallel execution of the model across the GPUs.
\textbf{Tensor parallelism (TP)} is the central mechanism for parallelizing model inference on multi-GPU systems. 
TP partitions the parameters \emph{and} computation of individual model layers across GPUs, so a single forward pass is executed collaboratively by multiple GPUs, improving throughput~\cite{shoeybi2020megatron}.
This way, the entire model need not be replicated on each GPU, thus overcoming the memory limitations of a single GPU faced by SSMs~\cite{deepspeedinference,kwon2023pagedattention}. 
Further, by splitting per-token compute across multiple GPUs, TP also helps to reduce per-request (or per-batch) latency.


Despite the prevalence and support of TP for other models, there is not yet a standard, ``drop-in'' TP implementation for SSMs and its variants.
%
For Transformers, for example, TP has converged to well-tested patterns because most of the work inside each Transformer block is a small number of large matrix multiplications, and it is relatively clear where to split those computations across GPUs~\cite{shoeybi2020megatron}. 
SSMs are organized differently. 
Each SSM consists of sequential SSM mixer blocks chained together to transform token embeddings into progressively richer representations, culminating in the final hidden states used to predict the next token.
SSMs also include a state-update operator that is applied across the sequence and lightweight local mixing such as convolution~\cite{gu2023mamba,dao2025transformersareSSMs}. 
These operations have \be{strict locality and memory-layout requirements} to achieve high throughput, and they interact with intermediate activations that are often larger than the final model-space output. 
As a result, applying Transformer TP templates, that are designed around sharding General Matrix-Matrix multiplications (GEMMs), can inadvertently place communication at unfavorable points in the mixer, leading to poor scaling.

In practice, designing an effective tensor-parallel SSM implementation is complicated by the following challenges:
\begin{itemize}[leftmargin=*]
  \item \textbf{Reusing per-layer state across phases.}
  Inference serving naturally has an initial prompt pass followed by token-by-token generation.
  This requires the internal state of the input prompt to be reprocessed for \emph{every} token generation, significantly increasing latency on the critical path.

  \item \textbf{Communication-aware sharding.}
  A good TP setup would shard work so each GPU can execute its portion of the mixer block using local tensors, without repeatedly reassembling intermediates.
  This requires choosing shard boundaries that match how the SSM block produces and consumes intermediate activations.
  Naively sharding on the first dimension of the local tensor (as is the case in successful Transformer TP implementations) leads to extra synchronization steps.

  \item \textbf{Handling packed parameter blocks.}
  SSM mixers often represent multiple logical quantities in a single packed tensor for efficiency.
  Uniformly slicing such packed tensors across GPUs can separate fields that must be available together, forcing expensive reconstruction steps (and hence extra communication).

  \item \textbf{Minimizing unavoidable inter-GPU communication.}
  While some cross-GPU communication is necessary to produce the correct block output under TP, it still imposes significant latency that affects end-to-end throughput.
\end{itemize}


Prior work has largely focused on algorithmic and kernel efficiency for SSMs on a \emph{single GPU}~\cite{gu2021s4,gu2023mamba,dao2025transformersareSSMs}. 
There are also early efforts to integrate specific SSM implementations into distributed training stacks on specialized platforms (e.g., Mamba-2 training with neuronx-distributed on AWS Trainium), underscoring practical demand for parallel execution~\cite{awslabsneuronssm}. 
However, to the best of our knowledge, \be{tensor-parallel inference for modern SSMs is not yet a well-established capability} in mainstream serving stacks as existing TP implementations for LLMs do not extend to SSMs~\cite{shoeybi2020megatron,deepspeedinference}.

In this paper, we present a tensor-parallel inference design for selective SSMs that directly addresses the above-mentioned practical challenges that arise in serving. 
The key contributions of our design are as follows:
\begin{enumerate}[leftmargin=*]
    \item We introduce a \be{distributed caching mechanism} that prevents reprocessing the input prompt for every newly-generated token by persisting the model’s per-layer internal state.
    The cache contents are carefully selected and sharded across GPUs to retain the locality needed for  continued computation without additional inter-GPU communication.
    \item We design an \be{intelligent tensor-parallel sharding scheme} that aligns shards with the mixer’s computation so that the core SSM-path operations stay GPU-local, avoiding communication introduced by naive splitting.
    \item We \be{handle the mixer’s packed parameters explicitly} by partitioning the packed block to preserve the  fields required for correct state updates, ensuring each GPU has the components it needs for its local update while distributing the remaining storage and compute across GPUs.
    \item We reduce the communication overhead of the TP synchronization step by \be{introducing quantized communication} tailored to the SSM's output aggregation during inference. 
\end{enumerate}



We experimentally evaluate our TP design on four representative SSM-based LLMs that cover both pure-SSM and hybrid architectures: Mamba, Mamba-2, Falcon-Mamba, and Zamba~\cite{gu2023mamba,dao2025transformersareSSMs, zuo2024falconmamba,glorioso2024zamba}. 
Tested on two GPU clusters, NVIDIA A6000 and NVIDIA A100, we show that our efficient TP design \be{increases throughput by 1.5--1.9$\times$ on 2 GPUs} and by \be{2.4--3.9$\times$ on 4 GPUs} compared to single-GPU inference, across all four SSMs, with the largest gains at long contexts. 
Importantly, our TP implementation allows us to support \be{significantly larger prompts sizes (2--4$\times$)} compared to single-GPU and data-parallel inference.
Furthermore, applying quantized AllReduce for slightly reduced precision improves throughput by an additional 10--18\% by lowering synchronization bandwidth overhead.


The rest of this paper is organized as follows. Section~\ref{sec:background} provides background on SSMs and tensor-parallel inference. Section~\ref{sec:Design} presents our key contribution: the tensor-parallel design for SSM inference. Section~\ref{sec:evaluation} evaluates performance and scaling behavior of our tensor-parallel design for four SSMs under two different GPU clusters.
We conclude this paper in Section~\ref{s:conc}.

\section{Background}
\label{sec:background}
In this section, we first discuss State Space Models (SSMs) in more detail, and then provide the necessary background on tensor parallelism.
\vspace{-4pt}
\subsection{Modern usage of SSMs}
\label{ss:bg_ssms}
\vspace{-4pt}
As SSMs have matured, they have become a \be{practical alternative to attention modules} for long-context modeling.
Compared to self-attention, which forms token-to-token interactions and therefore incurs \emph{quadratic} work in sequence length and a growing KV-cache during decoding, SSM mixers update a compact recurrent state in a streaming manner in \be{linear} time complexity.
As a result, SSMs can offer better long-context efficiency and lower decode-time memory overhead, which is especially attractive under serving constraints.

\textbf{Mamba} is a prominent selective SSM mixer that enables competitive attention-free language models by combining token-wise projections with efficient state-update computation~\cite{gu2023mamba}.
Building on this direction, \textbf{Falcon-Mamba} demonstrates a \emph{pure} Mamba-based, attention-free long-sequence generation of SSMs through substantially lower memory overhead~\cite{zuo2024falconmamba}.
In parallel, hybrid architectures have incorporated attention sparingly to recover retrieval and in-context learning behavior while retaining the efficiency of SSM backbones~\cite{falconH1}. 
\textbf{Zamba} proposes a compact hybrid model that pairs a Mamba backbone with a single shared self-attention module, aiming to preserve attention’s benefits at minimal parameter cost~\cite{glorioso2024zamba}.
\textbf{Mamba-2} introduces State Space Duality that reformulates the Mamba mixer’s state-update computation to map cleanly onto matrix-multiplication style kernels, easing scalability~\cite{dao2025transformersareSSMs}.
Together, the development of these model families indicates that SSMs are becoming a standard component of the LLM design space.

\vspace{-4pt}
\subsection{Overview of SSM architecture}
\label{ss:bg_ssm_arch}
\vspace{-4pt}
\begin{figure}[!t]
    \vspace{-20pt}
  \centering
  \includegraphics[width=0.95\columnwidth, height=3.8in]{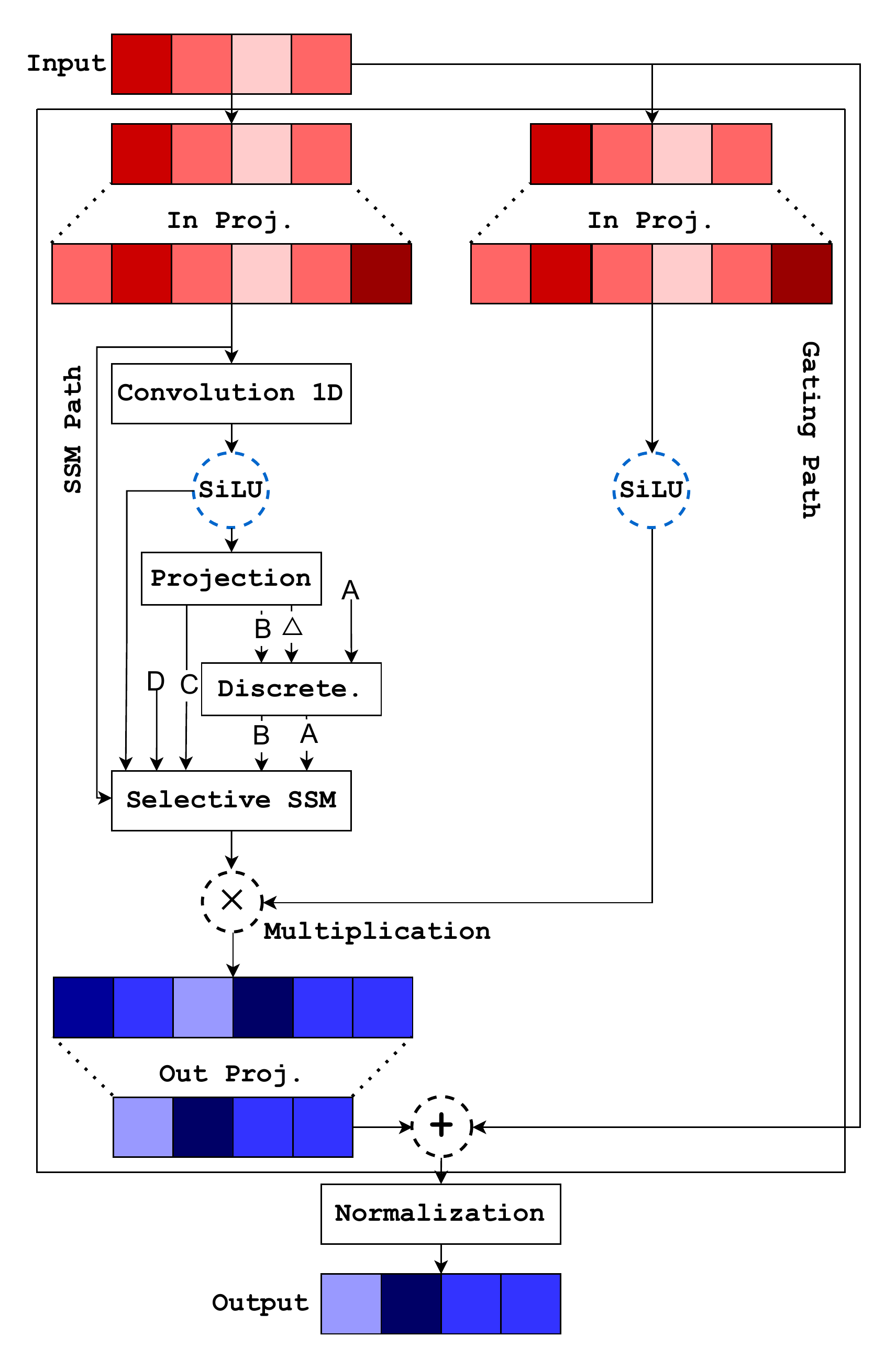}
  \caption{Mamba-style SSM mixer block. The input projection produces a packed tensor that is split into an SSM path and a gating path. The SSM path applies a channel-separable 1D convolution, projects activations into token-dependent SSM fields ($\Delta$, $B$, $C$) together with per-channel parameters ($A$, $D$), performs the state update over the sequence, gates the result, and applies an output projection back to the residual stream.}
  \label{fig:mamba_mixer_block}
  \vspace{-17pt}
\end{figure}

A state-space model (SSM) processes a sequence by maintaining a small internal \emph{state} that summarizes prior context and is updated as new tokens arrive.
Instead of computing pairwise interactions between all tokens (as in self-attention), an SSM updates this state over time and produces an output for each token from the evolving state.
For practical deployment, this ``streaming'' structure is appealing because prompt processing \be{scales linearly} with sequence length and token-by-token generation.
Modern SSM layers package these ideas into a practical \textbf{SSM mixer block} that combines Input projections with a state-update path as shown in Figure~\ref{fig:mamba_mixer_block}.
We next describe the typical Mamba-style mixer block at a high level in the same order as it is executed.

The first SSM mixer block receives the prompt activations as the residual-stream tensor \texttt{hidden\_states}.
An \emph{input projection} expands this representation into a wider intermediate tensor and is immediately split into two branches:
(i) an \emph{SSM path} that performs sequence modeling via convolution and a state update, and
(ii) a \emph{gating path} that produces a multiplicative gate used later to modulate the SSM-path output.

In the SSM path, a short 1D convolution is applied.
%
Next, a lightweight projection maps the convolved activations into the \emph{token-dependent parameter fields} used by the SSM update.
These fields are produced in packed form and then split into the logical quantities that drive the update:
\begin{itemize}[leftmargin=*,nosep]
  \item $\Delta$: a per-token \emph{step size} that controls how the continuous dynamics are converted into a discrete update.
  \item $B$ and $C$: per-token \emph{input/output} fields; $B$ determines how the current token influences the state update, and $C$ determines how the state is read out into an output for that token.
\end{itemize}
In addition to these token-dependent fields, the block uses learned per-channel parameters that are fixed for the layer:
\begin{itemize}[leftmargin=*,nosep]
  \item $A$: per-channel \emph{state dynamics} parameters that define how the state evolves over time.
  \item $D$: a per-channel \emph{skip/scale} term that provides a direct contribution from the activation stream to the output (often viewed as a residual-like bypass inside the mixer).
\end{itemize}

Before applying the recurrence, the block performs a \emph{discretization} step that combines $\Delta$ with the fixed dynamics $A$ (and scales $B$ accordingly) to form the coefficients used by the discrete-time update.
The \emph{SSM update} (also called the scan/state-update) then walks forward over the sequence: it updates the per-channel recurrent state at each token and emits a per-token output via $C$, with an additional per-channel contribution via $D$.
In practice, these token-dependent SSM quantities are generated as a single \emph{packed parameter tensor} and then \texttt{split} into $\Delta$, $B$, and $C$ for the state-update computation.

The SSM-path output is then combined with the gating path by elementwise multiplication (the gate controls how much of the SSM output is passed through).
Finally, an \emph{output projection} maps the result back to the model hidden size, and the block returns to the residual stream by adding this output to the incoming \texttt{hidden\_states}.

The same mixer structure also appears in recent model families that use Mamba-style SSM blocks as core building units, such as Mamba-2 Falcon-Mamba and Zamba.
From a systems and tensor-parallel perspective, both models still inherit the same key parallelization constraints for the SSM mixer path: intermediate activations are packed and then split into the fields needed by convolution and the SSM update.

\vspace{-4pt}
\subsection{Tensor parallelism and SSMs}
\vspace{-4pt}

Tensor parallelism (TP) is a model-parallel strategy that splits the parameters and computation of a \emph{single layer} across multiple GPUs so they collaboratively execute one forward pass~\cite{shoeybi2020megatron}.
In Transformer models, TP is commonly realized by sharding the large projection matrices in attention and linear layers and inserting a small number of communication collectives (e.g., AllGather/AllReduce) to form the correct activations.

The operator mix in modern SSM blocks differs materially from standard Transformer blocks. 
While SSM layers still include token-wise matrix multiplications that resemble Transformer linear projections, they also include sequence-wise kernels (selective scan/state update) whose data dependencies and intermediate-state structure are not identical to attention KV-cache access patterns. 
Because the SSM mixer’s state-update path requires per-channel contiguous intermediate slices (from packed parameter tensors) and local recurrent state updates, naively applying standard Transformer TP templates can fragment these layouts and introduce extra communication on the critical path.
Consequently, the TP strategies that work well for Transformers do not necessarily transfer directly to SSMs, and \be{must be revisited} with respect to SSM-specific computation graphs and state layouts. 
This gap motivates this work on SSM-aware TP design and implementation.

\section{Related Work}
\label{sec:related_work}
Prior work on SSM development has proposed hardware and compiler optimizations, including PIM-based acceleration and fine-grained GPU scheduling approaches~\cite{pimba,pipethreader}.
There is also work on algorithmic and kernel efficiency improvements for SSMs~\cite{gu2021s4,gu2023mamba,dao2025transformersareSSMs,zuo2024falconmamba,glorioso2024zamba}. 
However, all of these works are limited to a \emph{single-GPU} SSM deployment.
Consequently, the parallelization of SSMs remains underexplored despite the growing popularity of SSM-based model families.
%
To our knowledge, there is no published work on parallel implementations of SSMs. However, there is related work on parallelizing Transformer models and reducing the overheads of parallelization in multi-GPU settings, which are relevant to our design.
\vspace{-4pt}
\subsection{Transformer-oriented Tensor Parallelism (TP) techniques}
\label{ss:related-transformer}
\vspace{-4pt}

We discuss Transformer-oriented TP because Transformers have been the dominant sequential token-generation workload in practice, and most mature multi-GPU inference parallelism templates and systems were developed around Transformer blocks—making them the natural starting point (and baseline) when extending TP support to SSM mixers.
TP is well established for Transformer models, where compute is dominated by dense matrix multiplications and a small set of communication collectives.
Megatron-LM introduced practical column/row-parallel templates for Transformer layers and demonstrated scalable training and inference under TP configurations~\cite{shoeybi2020megatron,megatronpipeline}.
DeepSpeed and Colossal-AI provide end-to-end system stacks that integrate TP with memory optimizations, pipeline execution, and distributed runtime support~\cite{deepspeedinference,colossalai}.
These systems heavily influenced today’s production TP practices. 
However, as discussed in Section~\ref{ss:bg_ssm_arch}, SSM blocks introduce additional projections and state-space operations that require revisiting sharding and communication placement beyond the Transformer case~\cite{gu2023mamba,dao2025transformersareSSMs}.

\vspace{-4pt}
\subsection{Communication compression and quantization}
\label{ss:related-quantization}
\vspace{-4pt}
A key performance issue with multi-GPU TP designs is the communication latency that is incurred during inter-GPU communication.
In general, communication-efficient distributed training and inference has a long history of using compression and quantization techniques to reduce the payload of collective (communication) operations.
Communication overhead reduction techniques using low-bit/1-bit and stochastic quantization, as well as low-rank/structured compression, have been revisited by recent studies for optimizing large-scale model parallelism communication collectives~\cite{seide2014onebit,alistarh2017qsgd,vogels2019powersgd,xin2025global}.

At the systems level, modern GPU software stacks are also beginning to expose low-precision communication primitives (e.g., FP8 and FP16-oriented support in Transformer Engine), making quantization increasingly practical in end-to-end deployments~\cite{nvidiaTErelease25}.
Building on these techniques, our TP design for SSM inference includes an optional quantization component, as described in Section~\ref{sec:design:quantization}.

\section{Design}
\label{sec:Design}
\begin{figure*}[!t]
\vspace{-20pt}
  \centering
  \includegraphics[width=\textwidth]{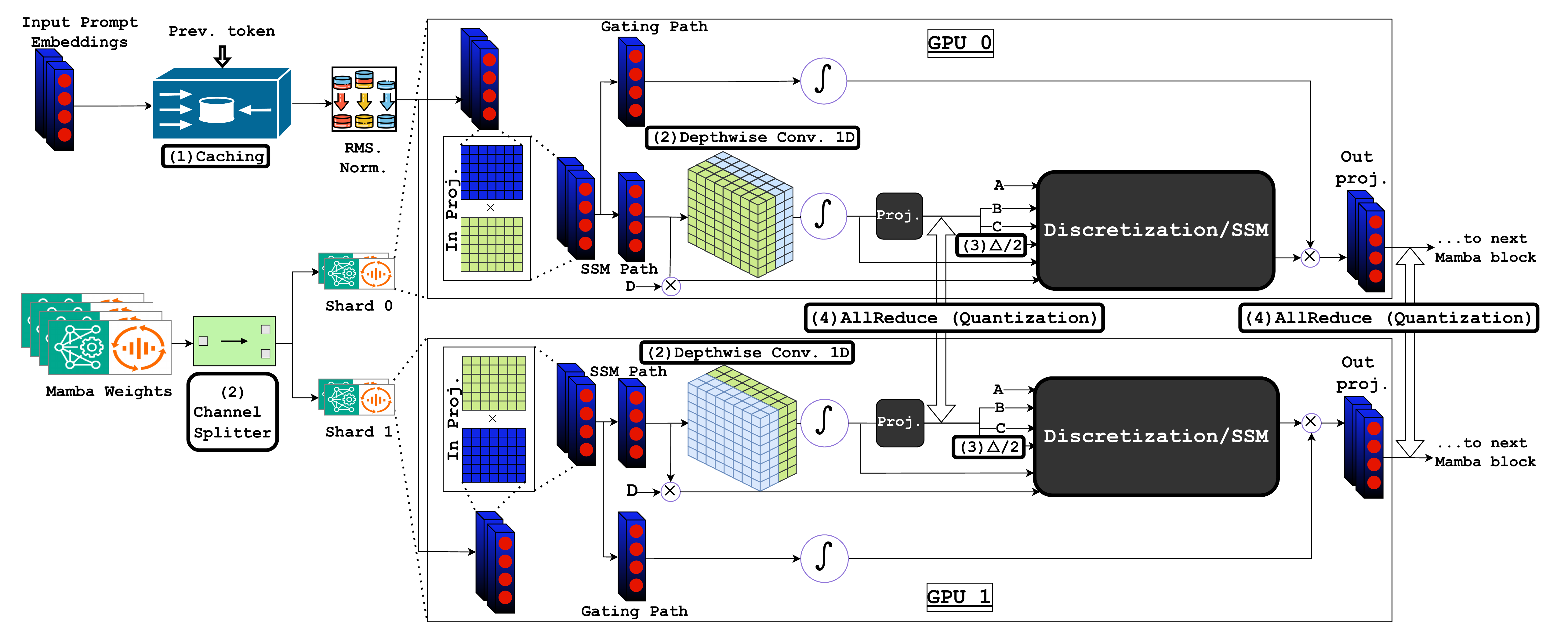}
  \caption{Illustration of our tensor-parallel inference implementation for Mamba, consisting of our four key design components: (1) SSM cache, (2) channel splitter and depthwise convolution, (3) packed parameter handling, and (4) AllReduce quantization.}
  \label{fig:design}
  \vspace{-17pt}
\end{figure*}

This section describes our key contribution: the system design for enabling \emph{tensor-parallel (TP) inference} for SSM-based models.
The core challenge is that an SSM mixer is not just a sequence of large matrix multiplications, as in a Transformer; it also contains locality-sensitive operators (notably depthwise convolution and an SSM state-update kernels) that achieve high single-GPU performance by operating on GPU-resident, contiguous tensors with minimal intermediate materialization.
A naive reuse of standard TP templates (such as those for Transformers) will \be{break locality}, thereby inducing  \emph{expensive and unnecessary inter-GPU communication} (AllGathers/AllReduces) and layout transforms at inappropriate locations, which increases memory traffic, breaks fused execution, and reduces throughput.

To make TP practical for SSMs, we make four key design contributions, in the following order, as shown in Figure~\ref{fig:design}:
(1) caching for low latency, 
(2) intelligent channel-wise model splitting to reduce synchronization and depthwise convolution, 
(3) handling of SSM parameter streams under sharding, and 
(4) optional AllReduce quantization.
We now discuss these design decisions in detail.
We commit to making our tensor-parallel SSM inference implementations publicly available, on acceptance.
\vspace{-4pt}
\subsection{Adding an SSM cache for low-latency serving}
\label{sec:design:caching}
\vspace{-4pt}
In autoregressive inference, including SSM-based inference, each new token is generated based on some \textbf{context}, which itself is a function of previously generated tokens.
So every time a token is to be generated, the context is reconstructed by processing the prior tokens.
As such, a key performance issue in SSM-based inference is the \be{repeated reconstruction of context} for each new token, which inflates end-to-end latency. 
This issue is further \emph{exacerbated under tensor parallelism (TP)} as the context must be replicated on each GPU,  \be{bloating its memory usage}~\cite{kwon2023pagedattention}.

To address this, we introduce an \be{SSM cache} (see Figure~\ref{fig:design}) that persists the minimal per-layer context needed to resume generation from the end of the prompt without rescanning the prefix.
With this SSM cache, the inference naturally decomposes into two phases: 
\emph{prefill}, which is a one-time phase that processes the full input prompt once to populate the cache; and 
\emph{decode}, which reads from the cache to get context from the previously generated tokens for each new token that needs to be generated. 
Essentially, at the expense of adding the prefill phase, the SSM cache significantly reduces the redundant processing time of prior tokens in each decode phase.

However, what to store in the SSM cache is not a trivial question under TP.
The reusable context in SSMs is \emph{not} a Transformer KV cache; instead it consists of (a) the compact per-layer SSM state produced after processing the prompt, and (b) a short convolution history used by the causal depthwise convolution.
Our SSM cache stores exactly these objects, and under TP it is \emph{sharded by channels} (hidden-feature dimensions) so that each GPU stores only the cache entries for the channels it owns.
This ensures \be{cache reads/writes stay GPU-local} and prevents the cache itself from adding extra inter-GPU synchronization during decoding to reconstruct the context.


\vspace{-4pt}
\subsection{Intelligent channel-wise splitting to minimize synchronization}
\label{sec:design:channel_splitting}
\vspace{-4pt}
A central bottleneck in TP is \be{inter-GPU communication}: if weight sharding (see Section~\ref{ss:bg_ssm_arch}) is misaligned with how the model packs and consumes activations, the execution repeatedly incurs extra AllReduces just to reassemble intermediate tensors, turning the critical path into a sequence of synchronization points. 
This is particularly harmful for SSM mixers, where the forward path relies on keeping intermediate layouts contiguous and GPU-local; improper sharding can therefore \emph{force additional layout transforms} which need to be synced, increasing communication volume and reducing effective throughput.

With naive sharding, the number of communication collectives can grow to four per block because packed intermediate tensors must be repeatedly reassembled.
Concretely, extra synchronization can be triggered (i) after the \emph{input projection} (to reconstruct the packed activation before it can be chunked), (ii) around the \emph{convolution branch} (when the sharded layout no longer matches the depthwise-convolution channel grouping), (iii) after the \emph{SSM-parameter projection} (to make the per-token parameter vectors consistent across ranks), and (iv) when \emph{merging} the SSM output with the gating branch (to re-form the full residual-stream hidden-state tensor).
See Section~\ref{ss:bg_ssm_arch} for a general description of the SSM architecture.

To address this, our TP design employs a \textbf{channel splitter} (see Figure~\ref{fig:design})
that \be{shards the weights along channels}, so each GPU owns a disjoint subset of channels and can run the mixer’s locality-sensitive operators on its local shard. 
We apply split-aware channel sharding inside the SSM mixer block: each GPU owns a disjoint subset of channels, and the channel splitter shards the mixer’s projection weights so that each rank produces the contiguous channel slices needed by its local downstream computation (depthwise convolution and the SSM update path).

Our channel splitter \be{lowers the required AllReduces from four (under naive sharding) to two} per block.
First, we perform an AllReduce on the mixer’s \emph{SSM-parameter projection}
This step ensures every GPU has the \emph{complete} parameter vectors needed for its local gating/SSM computation, without extra reconstruction collectives later.
Second, we perform an AllReduce at the \emph{residual-stream boundary}
(the end-of-block handoff point where the block’s output hidden-state tensor is passed to the next layer).
The residual-stream hidden representation is the main activation tensor that flows from block to block (the model’s \texttt{hidden\_states}).


To keep the mixer block's downstream operations local, we \be{implement
the mixer block’s 1D convolution to be channel-separable} (\texttt{Conv1d} on \texttt{groups} $=$ \texttt{intermediate\_size}), i.e., it mixes only across \emph{time} within each channel and never across channels in a single GPU.
As a result, when channels are sharded across GPUs, each rank can run the convolution (and the subsequent SSM scan/update) on its local contiguous channel shard without introducing communication in these paths.
Because the convolution is applied independently per channel and only slides across time, sharding channels across GPUs preserves correctness without introducing any communication in the convolution step.

\vspace{-4pt}
\subsection{Handling SSM parameter streams under TP}
\label{sec:design:ssm_params}
\vspace{-4pt}

Beyond projections and convolution, the mixer’s SSM update path consumes a set of SSM-related parameters that are produced/packed for efficient execution by the fused SSM kernel.
At a high level, these parameters define the per-token state update: $\Delta$ controls the input-dependent discretization step (i.e., how the continuous-time dynamics are converted into a per-token update), $A$ governs the recurrent dynamics of the state, $B$ determines how the current token drives the state update, $C$ maps the state back to an output contribution, and $D$ provides a direct residual/skip contribution to the output~\cite{gu2023mamba,dao2025transformersareSSMs}.
However, these parameters do not all behave the same way, so treating them as one uniformly sharded tensor can be both incorrect and inefficient.

In Mamba-style implementations, $A$ and $D$ are learned \emph{per-channel} parameters, while $\Delta$, $B$, and $C$ are \emph{token-dependent} quantities produced from activations (implementation details vary across codebases, but the practical systems requirement is consistent).
Operationally, however, common implementations pack these quantities into a single \texttt{SSM\_parameters} tensor, where different column ranges correspond to $\Delta$, $B$, $C$ (and related packed fields), and the fused SSM kernel expects this packed layout.

This packing is convenient on a single GPU, but it is not TP-friendly because the packed tensor implicitly assumes a fixed contiguous layout, whereas TP needs a partitioning that depends on the chosen TP degree.
If we TP-shard the packed \texttt{SSM\_parameters} tensor naively, each rank can receive partial slices of multiple logical fields, \be{forcing either time-consuming reassembly collectives or extra layout transforms} before invoking the fused kernel.

To avoid this, we \be{explicitly unpack the logical fields} required by the fused SSM kernel and apply a TP-aware placement that keeps the state update GPU-local.
Specifically, we \emph{shard $\Delta$ with channels} (it is token-dependent, the largest stream, and dominates activation-side memory), while ensuring that all remaining quantities required to advance the SSM state for the owned channels are \emph{locally available} on every rank when the fused kernel runs.
In our implementation, this means token-dependent terms such as $B$ and $C$ are produced locally from the local activations, while per-channel learned parameters such as $A$ and $D$ are replicated (or stored) locally as needed.
This layout allows each rank to execute the SSM update using the same fused kernel on \emph{contiguous local shards} without any communication inside the SSM path, and it eliminates an otherwise required collective that would be needed to reconstruct packed parameter slices prior to the SSM operation.
\vspace{-4pt}
\subsection{Quantized AllReduce for the remaining TP synchronizations}
\label{sec:design:quantization}
\vspace{-4pt}

After the above design changes, TP communication is confined to a small number of mandatory AllReduces.
To further reduce the cost of these remaining yet expensive AllReduce collectives, we optionally apply \be{AllReduce quantization}; see Figure~\ref{fig:design}.
Quantizing the AllReduce payload is an intuitive optimization: reducing the communicated precision decreases bytes transferred per collective and therefore the AllReduce latency. 
Prior work on generic quantization has shown that aggressive low-precision exchange can preserve model quality (often with negligible accuracy loss) while substantially reducing communication overhead~\cite{seide2014onebit,alistarh2017qsgd,vogels2019powersgd}.

We implement the AllReduce quantization by only quantizing the communicated tensors to a lower-precision representation (FP16, from FP32) for transfer and reduction, and then dequantizing them back for subsequent computation.
With this selective quantization, we target the bottleneck boundary collectives, while leaving the locality-sensitive SSM and convolution kernels unchanged.
As shown in our evaluation (Section~\ref{ss:quantization}), our AllReduce quantization reduces bandwidth demand and improves end-to-end throughput. 
\vspace{-4pt}
\subsection{Extending TP to Mamba-2, Falcon-Mamba and Zamba}
\vspace{-4pt}

Our design extends directly to other Mamba-style model families with the same mixer structure.
\textbf{Falcon-Mamba} uses the same selective SSM mixer as its backbone, so we apply the same cache layout, channel-wise splitting, and packed-parameter handling without changing the core design~\cite{zuo2024falconmamba}.
\textbf{Zamba} retains Mamba-style mixers for most layers and adds a lightweight attention component, so we apply our TP design to the mixer layers while leaving the attention component to standard column splits TP where applicable~\cite{glorioso2024zamba}.
\textbf{Mamba-2} preserves the same high-level mixer pipeline (projection, convolution, state update, gating, output projection), so we reuse the same sharding and communication principles and only adapt to the specific packed parameter layout used by its implementation~\cite{dao2025transformersareSSMs}.


\section{Evaluation}
\label{sec:evaluation}
We now present our experimental evaluation results highlighting the performance gains afforded by our tensor-parallel (TP) inference design for SSMs. 
We start in Section~\ref{ss:esetup} by describing our experimental setup, SSM models employed, the baselines for comparison, and the experimental methodology.
We then present our key performance evaluation results on the increased prompt lengths enabled by our TP design (Section~\ref{ss:sizes}) and its throughput gains (Section~\ref{ss:results}), including with quantization (Section~\ref{ss:quantization}), and an ablation study (Section~\ref{ss:ablation}) to highlight how our TP design contributes to the observed performance improvements.
\vspace{-4pt}
\subsection{Experimental setup and methodology}
\label{ss:esetup}
\vspace{-4pt}

We evaluate our TP design on two multi-GPU platforms, representative of modern AI serving infrastructure.

\begin{enumerate}[leftmargin=*]
    \item \textbf{A6000 PCIe workstation.} A single server with an AMD EPYC Milan 7543P CPU (32 cores) and four NVIDIA RTX A6000 GPUs and PCIe~4.0 interconnect.
    \item \textbf{A100 NVLINK cluster.} A multi-user cluster (queue \texttt{a100}) with AMD Milan CPU (96 cores), four NVIDIA A100 GPUs, and NVLINK interconnect.
\end{enumerate}

\vspace{-10pt}
\medskip\noindent\emph{Models:}
We evaluate four SSM-based LLMs: \textbf{Mamba}~\cite{gu2023mamba}, \textbf{Mamba-2}~\cite{dao2025transformersareSSMs}, \textbf{Falcon-Mamba}~\cite{zuo2024falconmamba}, and \textbf{Zamba}~\cite{glorioso2024zamba}.
We chose these models to span diverse SSM implementations and pure-SSM versus hybrid designs, enabling a representative evaluation of our tensor-parallel design principles. 
Because \textbf{Mamba} and \textbf{Mamba-2} use a higher embedding (hidden) dimension than the other models, we evaluate them up to input/output sequence lengths of 256, whereas we evaluate \textbf{Falcon-Mamba} and \textbf{Zamba} up to 1024.



\vspace{-6pt}
\medskip\noindent\emph{Comparison baselines:}
Our TP design shards the model so that multiple GPUs collaboratively execute each request.
Our SSM-aware sharding, as described in Section~\ref{sec:Design}, keeps the recurrent update and cached state local while minimizing synchronization overhead in the mixer. For evaluation, we compare our TP design against two inference serving strategies.

\begin{itemize}[leftmargin=*]
    \item \textbf{Default (1$\times$)} runs the model on a single GPU with no model parallelism. This represents the current, de facto practice.
    \item \textbf{Data parallelism (DP)} replicates the full model across GPUs and distributes requests across replicas. DP is a competitive baseline for throughput, especially at higher request concurrency, but it does not reduce the latency (or work done) per request.
\end{itemize}

\begin{figure}[!t]
  \centering
  \includegraphics[width=\linewidth]{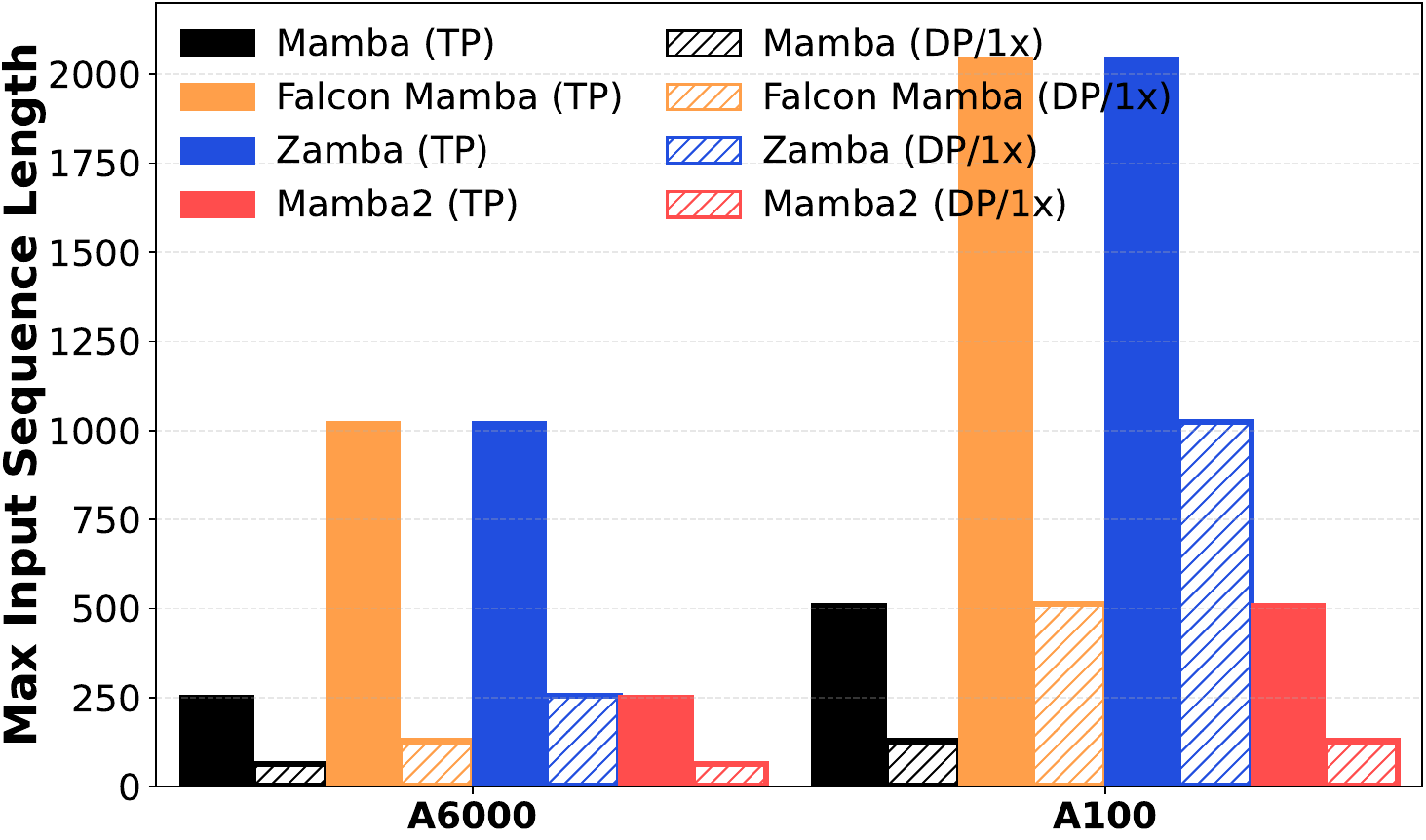}
  \caption{Maximum input sequence length possible at fixed (256) batch size under our TP design and under DP and 1x.}
  \label{fig:sizes}
  \vspace{-15pt}
\end{figure}
\vspace{-8pt}
\medskip\noindent\emph{Experimental methodology:}
For both platforms (A6000 and A100), we run 2-GPU and 4-GPU configurations and sweep across prompt (input) lengths, $L_{\text{in}}$, and generation (output) lengths, $L_{\text{out}}$, to cover short-, medium-, and long-context regimes for all models. Unless stated otherwise, all methods (our TP design and the two comparison baselines) use the same model weights, numerical precision, kernels, and decoding strategy.
All experiments use the Simple English Wikipedia (SimpleWiki) corpus as the evaluation dataset~\cite{coster2011simplewiki}.

\vspace{-4pt}
\subsection{Results on enabling longer prompt sizes}
\label{ss:sizes}
\vspace{-4pt}

A key benefit of our TP design is enabling larger prompt sizes ($L_{\text{in}}$, or input sequence length) for a given GPU memory by parallelizing the model across GPUs. 
To evaluate this benefit, we fix the batch size at 256, and increase the input sequence length in powers of 2 for each method until we encounter an OOM (Out-Of-Memory) error.
Figure~\ref{fig:sizes} shows maximum successful input length under each method for all models on both GPUs; note that DP and no parallelism (1$\times$) host the entire model on a GPU, and so have the same maximum input length.

We see that our TP design \be{consistently supports much longer input prompts} than DP or 1x on both A6000 and A100.
For Mamba and Mamba-2, our TP design supports 4$\times$ higher input lengths compared to DP/1x on both GPU architectures (256 vs.\ 64 on A6000 and 512 vs.\ 128 on A100).
Likewise, for Falcon-Mamba and Zamba, we support 2--4$\times$ higher input lengths.
This behavior is expected because our TP design shards model weights and per-layer runtime state across GPUs, reducing the per-device memory footprint.
By contrast, DP replicates the full model on every GPU, so each replica hits the same memory ceiling as 1$\times$ and cannot admit longer prompts even when more GPUs are available.
We note that the memory efficiency enabled by our TP design can also translate to supporting large SSM sizes on a given GPU.

\begin{figure*}[!ht]
    \vspace{-15pt}
    \centering
    \includegraphics[width=\linewidth]{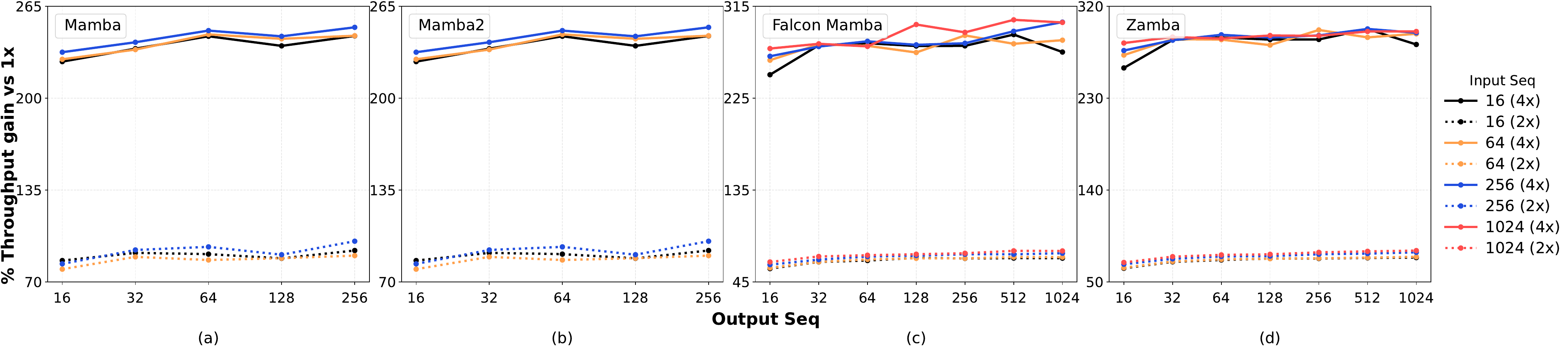}\\[9pt]  
    \includegraphics[width=\linewidth]{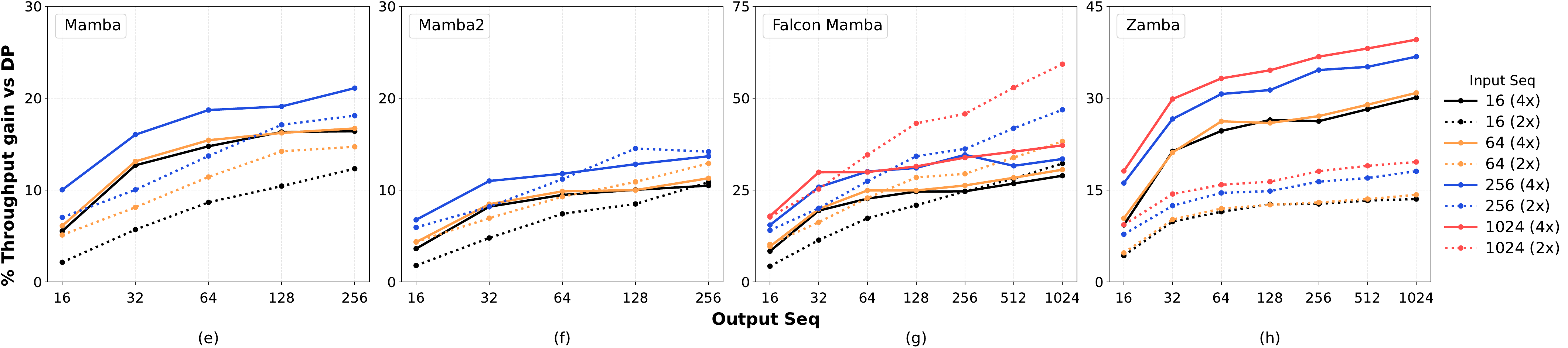}
    \vspace{-0.5cm}
    \caption{Throughput gains afforded by our tensor-parallel inference design (for 2-GPU, 4-GPU) for Mamba, Mamba-2, Falcon-Mamba, and Zamba, compared to no parallelism  (``1x'', first row) and compared to data parallelism (``DP'', second row) on the A6000 cluster.}
    \label{fig:tp_throughput_A6000}
    \vspace{-15pt}
\end{figure*}

\vspace{-4pt}
\subsection{Results on throughput gains}
\label{ss:results}
\vspace{-4pt}

We now report the throughput gains afforded by our TP design.
For this, we run experiments in a throughput-oriented serving condition. Specifically, for each $\{L_{\text{in}}, L_{\text{out}}\}$ configuration, we \emph{maximize the batch size} separately for each method until the run reaches the memory limit. We do this to ensure a fair comparison of the best achievable throughput under each parallelization strategy.

\begin{figure*}[!t]
    \centering  
    \includegraphics[width=\linewidth]{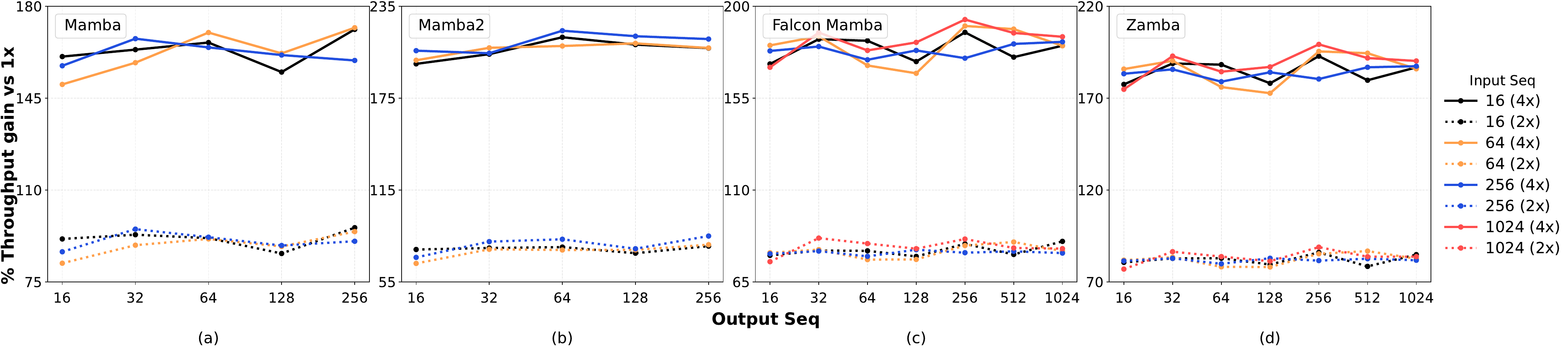}\\[9pt]
    \includegraphics[width=\linewidth]{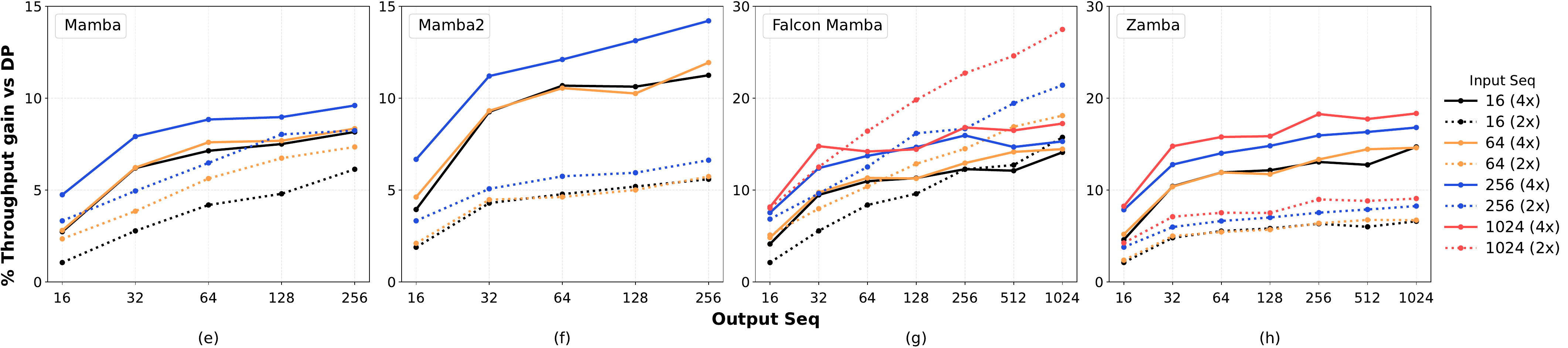}
    \vspace{-0.5cm}
    \caption{Throughput gains afforded by our tensor-parallel inference design (for 2-GPU, 4-GPU) for Mamba, Mamba-2, Falcon-Mamba, and Zamba, compared to no parallelism  (``1x'', first row) and compared to data parallelism (``DP'', second row) on the A100 cluster.}
    \label{fig:tp_throughput_A100}
    \vspace{-0.4cm}
\end{figure*}

Figures~\ref{fig:tp_throughput_A6000} and~\ref{fig:tp_throughput_A100} show the throughput improvement afforded by our tensor parallel (TP) design under the A6000 and the A100 GPU clusters, respectively.
In each figure, we show the results for each of the four models evaluated (Mamba, Mamba-2 Falcon-Mamba, and Zamba), shown as four columns of subfigures. 
The first row of subfigures shows the improvement compared to the default no parallelism (referred to as ``1x'') method and the second row of subfigures shows the improvement compared to data parallelism (DP).
In each subfigure, the $x$-axis is the output sequence length ($L_{\text{out}}$, number of generated tokens). 
As indicated by the legend, each curve in a subfigure corresponds to a fixed input prompt length ($L_{\text{in}}$) and the number of GPUs employed for parallelism (for DP and TP); we use solid lines to depict 4-GPU results and dotted lines to depict 2-GPU results.
\vspace{-4pt}
\medskip\subsubsection{Performance under NVIDIA A6000}
As shown in the first row of Figure~\ref{fig:tp_throughput_A6000}, our TP design provides significant performance improvement compared to the default, no-parallelism method, with \be{throughput gains of 58--98\% on 2 GPUs and 226--298\% on 4 GPUs} across the four models shown.
The throughput gain achieved by our TP design is primarily because of the scaling benefit it enables of splitting one request across multiple GPUs; while also accommodating larger batch sizes per inference. This is also why the gains are higher for 4 GPUs compared to 2 GPUs.
While not significant, the throughput gain for all models slightly increases with longer input/output lengths, indicating that our TP method is most beneficial in the long-context regimes that stress memory pressure.

Compared to DP, our TP design continues to provide substantial performance improvement, with \be{throughput gains of 2--59\% on 2 GPUs and 9--40\% on 4 GPUs} across the four models as shown in the second row of Figure~\ref{fig:tp_throughput_A6000}.
For Mamba (Figure~\ref{fig:tp_throughput_A6000}(e)), TP provides consistent gains that grow with the output sequence length and then mildly plateau, 
across the input-length sweep. 
For Mamba-2 (Figure~\ref{fig:tp_throughput_A6000}(f)), trends are similar as observed on Mamba
across the input-length sweep. 
Falcon-Mamba (Figure~\ref{fig:tp_throughput_A6000}(g)) shows larger gains and a stronger dependence on output sequence length;
the benefit is most pronounced at longer contexts where DP cannot reduce per-request work. 
Zamba (Figure~\ref{fig:tp_throughput_A6000}(h)) follows a similar pattern to Mamba but with higher gains 
and a gradual saturation with output length.

To understand the behavior of the throughput gains, we need to analyze how our TP design impacts the \emph{prefill} and \emph{decode} phases of the model; we discuss this in detail in Section~\ref{sss:ttft_tpot_behavior}.


\vspace{-4pt}
\medskip\subsubsection{Performance under NVIDIA A100}
The first row of Figure~\ref{fig:tp_throughput_A100} reports throughput gains over the default, no-parallelism baseline under the A100 cluster. Similar to the A6000 results, the dominant effect is the splitting of one request across multiple GPUs, so the curves are high and relatively flat across output lengths.
Compared to the A6000 results, the performance improvement afforded by our TP design on the A100 cluster is slightly lower, especially for the 4 GPU configuration, with \be{throughput gains of 75--96\% on 2 GPUs and 150--199\% on 4 GPUs} across the four models.
We observe larger TP-over-DP/1x gains on A6000 than on A100, since the A100’s higher compute and bandwidth strengthen the DP baseline and leave less headroom for model-parallel speedups.
The gains generally grow with longer input/output lengths before saturating, reinforcing that TP is most beneficial in long-context regimes where single-GPU execution becomes limiting.

The second row of Figure~\ref{fig:tp_throughput_A100} shows the throughput gains over DP. 
For Mamba (Figure~\ref{fig:tp_throughput_A100}(e)) and Mamba-2 (Figure~\ref{fig:tp_throughput_A100}(f)), TP provides modest but consistent gains that increase with output sequence length and then plateau, reaching roughly 3--15\% on 4 GPUs and 1--8\% on 2 GPUs across the input-length sweep. 
Falcon-Mamba (Figure~\ref{fig:tp_throughput_A100}(g)) shows larger gains with stronger dependence on output length, with TP improving throughput by about 4--17\% on 4 GPUs and 2--27\% on 2 GPUs as output length increases to 1024 tokens. 
Zamba (Figure~\ref{fig:tp_throughput_A100}(h)) follows a similar qualitative trend, achieving 4--16\% gains on 4 GPUs and 2--10\% gains on 2 GPUs, with gains generally rising with output length.
Overall, TP is expected to scale with GPU count (especially over NVLINK and, to a lesser extent, PCIe), but in rare cases such as Falcon-Mamba (Figure~\ref{fig:tp_throughput_A100}(g)) and (Figure~\ref{fig:tp_throughput_A6000}(g)) vs.\ DP, collective communication can dominate and make 4-GPU TP slower than 2-GPU TP.



\vspace{-4pt}
\medskip\subsubsection{Understanding the throughput gains}
\label{sss:ttft_tpot_behavior}
Our TP design impacts the model performance at both the prefill phase (processes the full input prompt once to populate the cache) and the decode phase (reads from the cache to get context from the previously generated tokens for each new token, see Section~\ref{sec:design:caching}). We measure these phases using two standard serving metrics: \textbf{time-to-first-token (TTFT)}, dominated by prefill plus any one-time setup until the first generated token is produced, and \textbf{time-per-output-token (TPOT)}, the steady-state per-token cost during decode.
\begin{figure}[!t]
    \vspace{-15pt}
  \centering
  \includegraphics[width=0.9\linewidth]{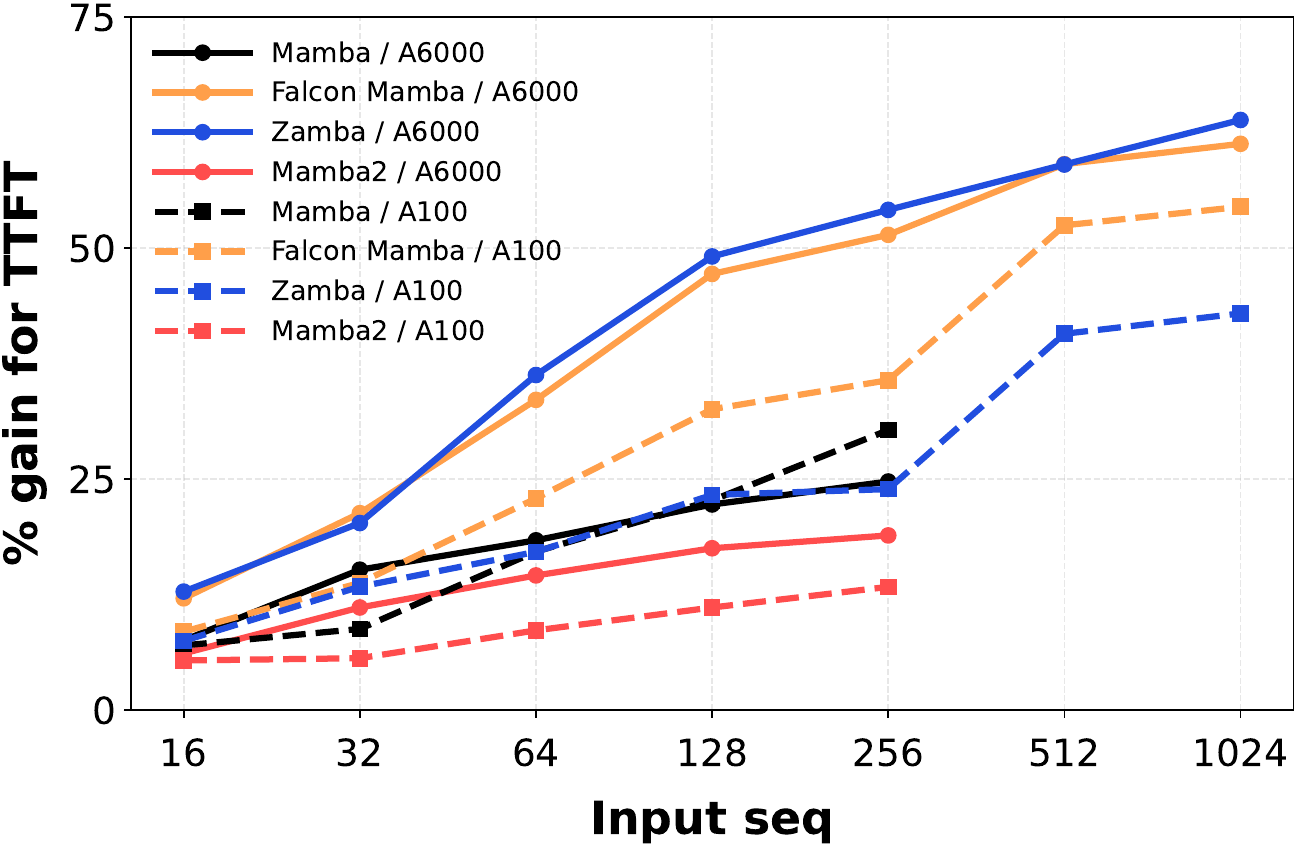}
  \caption{TTFT gain (vs.\ DP) afforded by our TP on 4 GPUs.}
  \label{fig:tp_ttft}
  \vspace{-15pt}
\end{figure}

To understand the impact, we now analyze the TTFT and TPOT metrics, that capture the latency of the prefill and decode phases, respectively.
We report TTFT and TPOT separately because they scale with different (input or output) sequence dimensions. 
Note that the throughput results in Figures~\ref{fig:tp_throughput_A6000} and \ref{fig:tp_throughput_A100} show the end-to-end throughput gains, which inherently include the contributions of TTFT and TPOT improvements.

Figure~\ref{fig:tp_ttft} shows the TTFT gains as a function of \emph{input sequence length} for 4-GPU TP relative to DP. 
Our measurements confirm that TTFT improvements are largely insensitive to the output sequence length, so we only plot TTFT gains for different input lengths. 
Across all models, TTFT gains increase monotonically with input (prompt) length, with the largest benefits appearing in the long-context regime where prefill dominates. 
On the A6000 (solid lines), Falcon-Mamba and Zamba scale from roughly 12--13\% at 16 tokens to about 61--64\% at 1024 tokens, but the gains slow down at longer prompts (e.g., only a small increase from 512 to 1024), indicating a mild plateauing effect as fixed overheads are amortized and the prefill path becomes increasingly bandwidth/compute limited. 
Mamba shows the same qualitative trend (rising to $\sim$25\% by 256 tokens); recall that Mamba encounters OOM error beyond 256 input length, so we only show Mamba results until then. 
On the A100 (dotted lines), gains are generally smaller at short prompts but remain substantial at long contexts: Falcon-Mamba gains $\sim$54\% and Zamba gains $\sim$43\% at 1024 tokens, again with diminishing returns toward the longest prompts, while Mamba achieves $\sim$30\% gains by 256 tokens.

\begin{figure}[!t]
  \vspace{-15pt}
  \centering
  \includegraphics[width=0.9\linewidth]{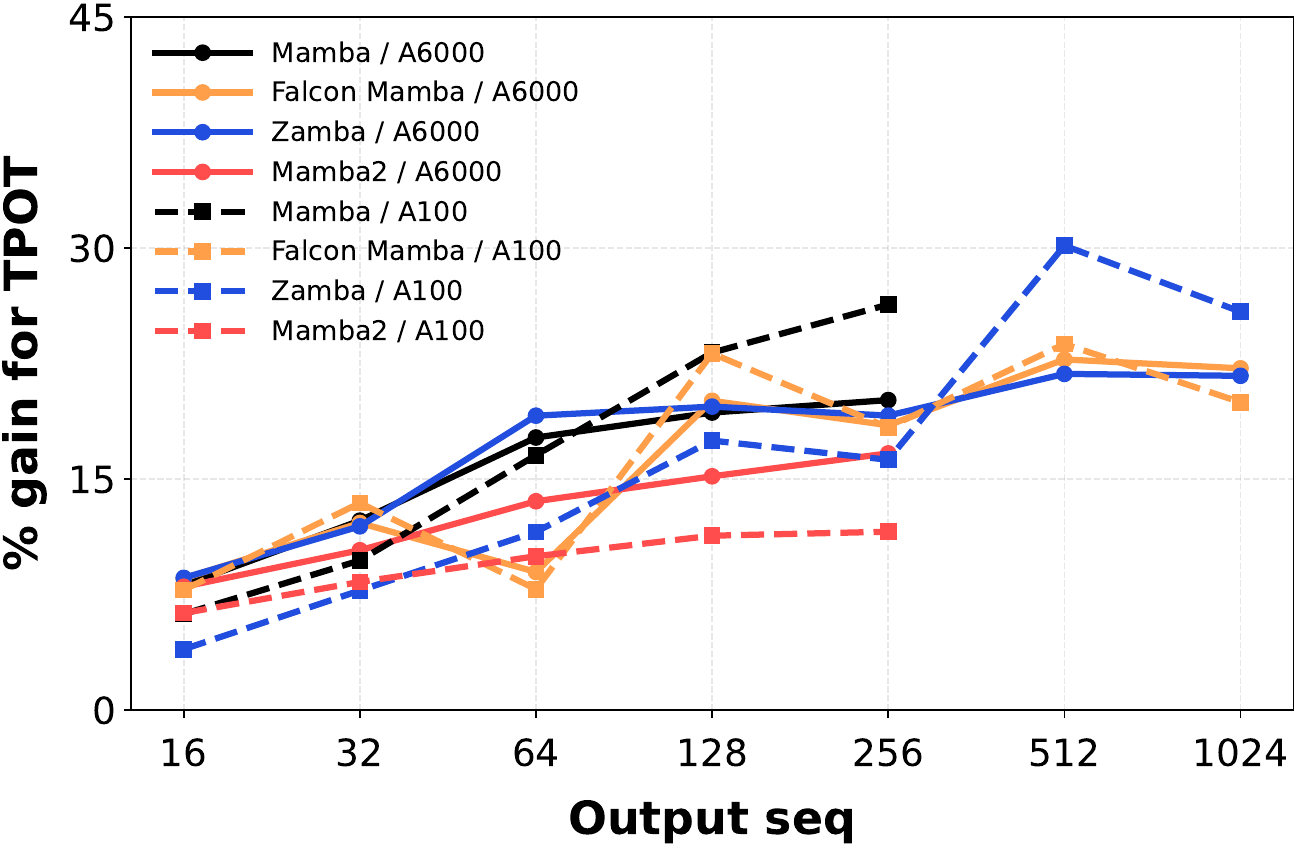}
  \caption{TPOT gain (vs.\ DP) afforded by our TP on 4 GPUs.}
  \label{fig:tp_tpot}
  \vspace{-15pt}
\end{figure}

Figure~\ref{fig:tp_tpot} plots TPOT gains as a function of \emph{output sequence length} for TP relative to the baseline. 
In contrast to TTFT, TPOT is much more strongly characterized by the output sequence length (decode length) than by the prompt length once the prefill state is established; as such, we only plot TPOT gains for different output lengths to capture steady-state decode behavior and how the per-token gains evolve as generation proceeds.
Across all models and both GPUs, we find that TPOT gains rise quickly from short decodes and then plateau as the output length increases: once the system reaches steady-state decoding, the per-token work and the placement of the required TP aggregation collectives are essentially constant, so additional generated tokens do not materially change the average time-per-output-token (TPOT). 

\noindent\textbf{Diminishing returns and utilization.}
Because TTFT (prefill) and TPOT (decode) exercise fundamentally different bottlenecks (TTFT is primarily compute/SM- and kernel-execution dominated, while TPOT is often dominated by memory footprint and bandwidth), the observed throughput gains under TP can be better understood by profiling the phase-appropriate resource usage (GPU compute utilization for TTFT and memory utilization for TPOT).

\begin{figure}[!t]
  \centering
  \begin{subfigure}[t]{0.495\columnwidth}
    \centering
    \includegraphics[width=\linewidth]{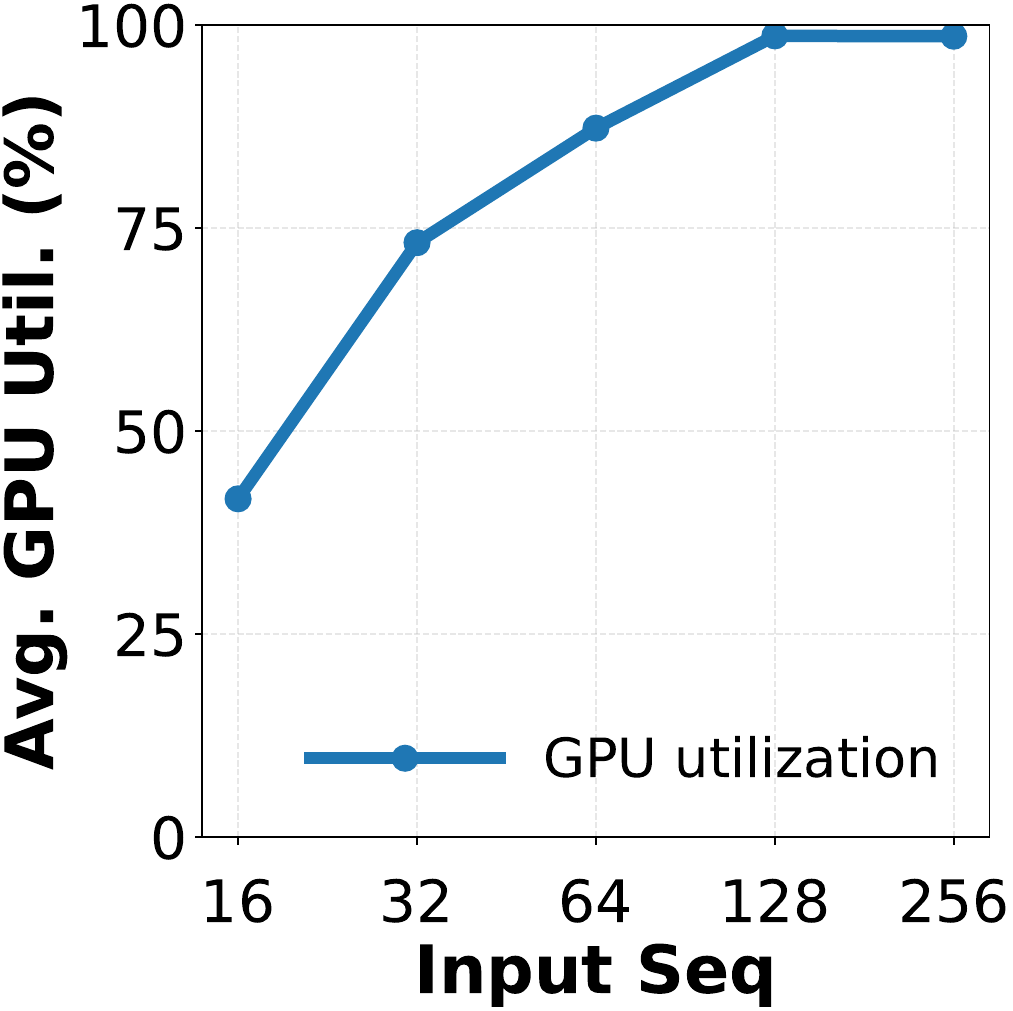}
    \caption{Average GPU utilization}
    \label{fig:gpu_util_vs_input_seq}
  \end{subfigure}\hfill
  \begin{subfigure}[t]{0.495\columnwidth}
    \centering
    \includegraphics[width=\linewidth]{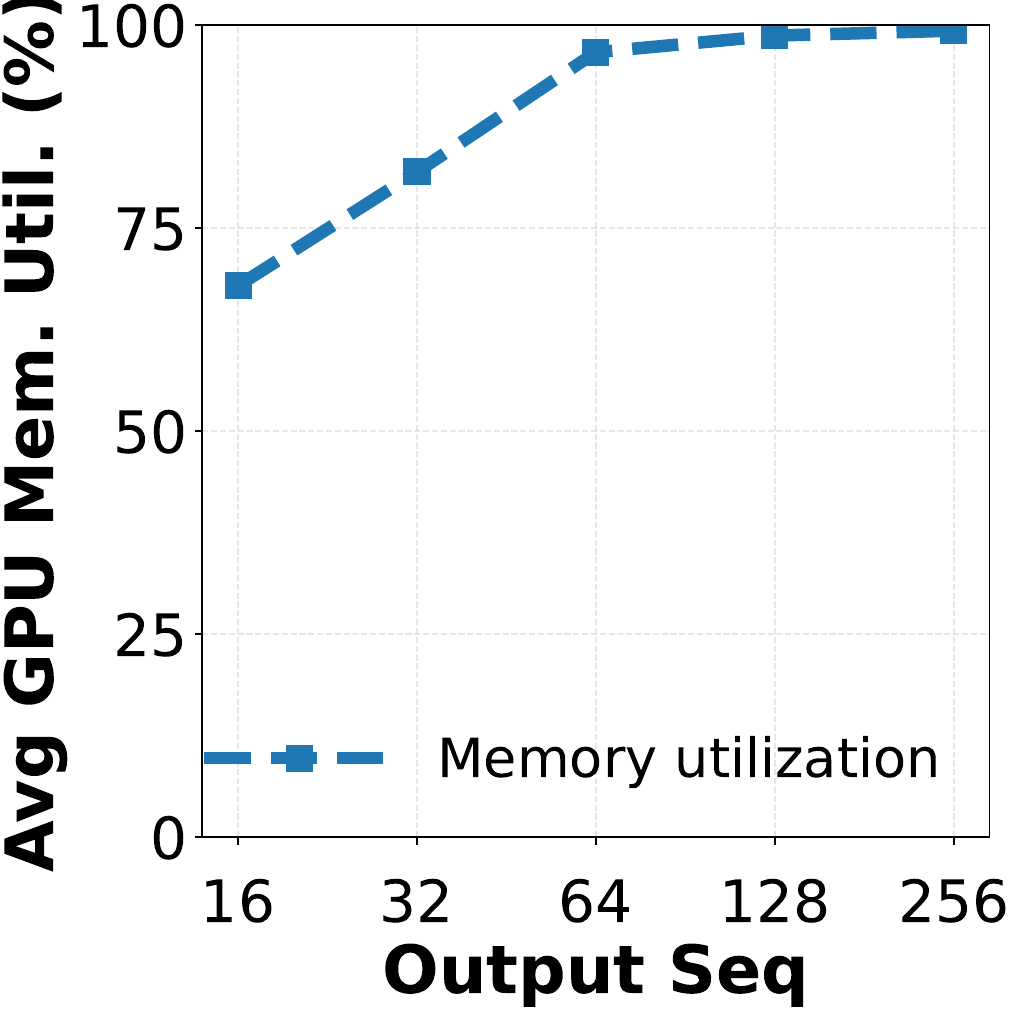}
    \caption{Average memory utilization}
    \label{fig:mem_util_vs_output_seq}
  \end{subfigure}
  \caption{GPU and memory utilization during TTFT and TPOT.}
  \label{fig:utilization_ttft_tpot}
  \vspace{-10pt}
\end{figure}

Figure~\ref{fig:gpu_util_vs_input_seq} shows the average GPU utilization as a function of input length during TTFT (prefill) for Mamba on NVIDIA A6000.
We see that the GPU utilization initially rises sharply with increasing input length,
but then saturates (at 98.68--98.64\% for 128--256 tokens). This indicates that prefill becomes compute-saturated at moderate-to-long prompts, leaving little headroom for TP to extract further throughput gains. 
Figure~\ref{fig:mem_util_vs_output_seq} shows the average GPU memory utilization as a function of output length during TPOT (steady-state decode).
We again see that memory utilization initially increases
and then saturates,
suggesting decode is increasingly constrained by memory capacity/bandwidth pressure from activations and recurrent/state buffers. 
In this regime, TP's remaining benefits are offset by the fixed cost of cross-GPU collectives and synchronization that TP execution entails, similar to the well-known source of diminishing returns in model-parallel language model systems~\cite{shoeybi2020megatron,kwon2023pagedattention}.

In summary, while TP increases throughput by enabling larger feasible batch sizes (via parameter/activation sharding) and by distributing the per-token compute across GPUs, its marginal benefit tapers as runs approach fundamental compute/memory limits and as fixed collective overheads become a larger fraction of each step. This behavior is consistent with classic strong-scaling limits: as the non-parallelizable and synchronization components grow in relative cost, speedups saturate (and can even regress) despite adding more parallel resources~\cite{williams2009roofline,amdahl1967validity}. 
This interpretation is consistent with the slight plateauing in Figures~\ref{fig:tp_throughput_A100} and~\ref{fig:tp_throughput_A6000} at larger output lengths.


\vspace{-4pt}
\subsection{Performance enhancement via AllReduce quantization}
\label{ss:quantization}
\vspace{-4pt}

While our TP design unlocks performance gains via parallel GPU execution, it does introduce additional inter-GPU communication; the results shown thus far include the impact of both these aspects.
To further improve the performance afforded by our TP design, we enable quantized AllReduce for tensor-parallel inference and measure its impact on both accuracy and performance; we  quantize from the default FP32 precision to FP16. 
We note that quantization does reduce model accuracy, so this enhancement essentially presents a tradeoff between accuracy and additional throughput gain.

\begin{table}[H] 
\centering
\small
\begin{tabular}{c|c|c}
\hline
\textbf{Model} & \textbf{Metric} & \textbf{Value (\%)} \\
\hline
Mamba & Top-1 (argmax) & 98.81 \\
Mamba & Top-5 overlap (Unordered) & 99.03 \\
Mamba & Top-5 (Ordered) & 89.01 \\ \hline

\hline
\end{tabular}
\caption{Accuracy impact of quantization for Mamba.}
\label{tab:mamba_quant_allreduce_accuracy}
\vspace{-10pt}
\end{table} 
We first quantify the accuracy impact by comparing model outputs against the non-quantized model using agreement-based metrics (Top-1 token match and Top-k overlap). 
Note that the GPU choice (A6000 or A100) does not impact the model accuracy.
Table~\ref{tab:mamba_quant_allreduce_accuracy} shows that our quantized AllReduce incurs only small output perturbations: the next-token prediction under quantization matches the non-quantized version $\sim$98\% of the time and the Top-5 candidate set overlaps $\sim$99\% of the time, suggesting that quantization rarely changes the model's preferred token and typically only perturbs the ranking among close-probability alternatives (reflected in the stricter Top-5 \emph{exact ordering} match of 87--89\% across models). These results align with the established view that low-precision communication is a favorable tradeoff when communication becomes a bottleneck~\cite{seide2014onebit,alistarh2017qsgd,vogels2019powersgd}.

\begin{figure}[t]
  \centering
  \includegraphics[width=0.9\linewidth]{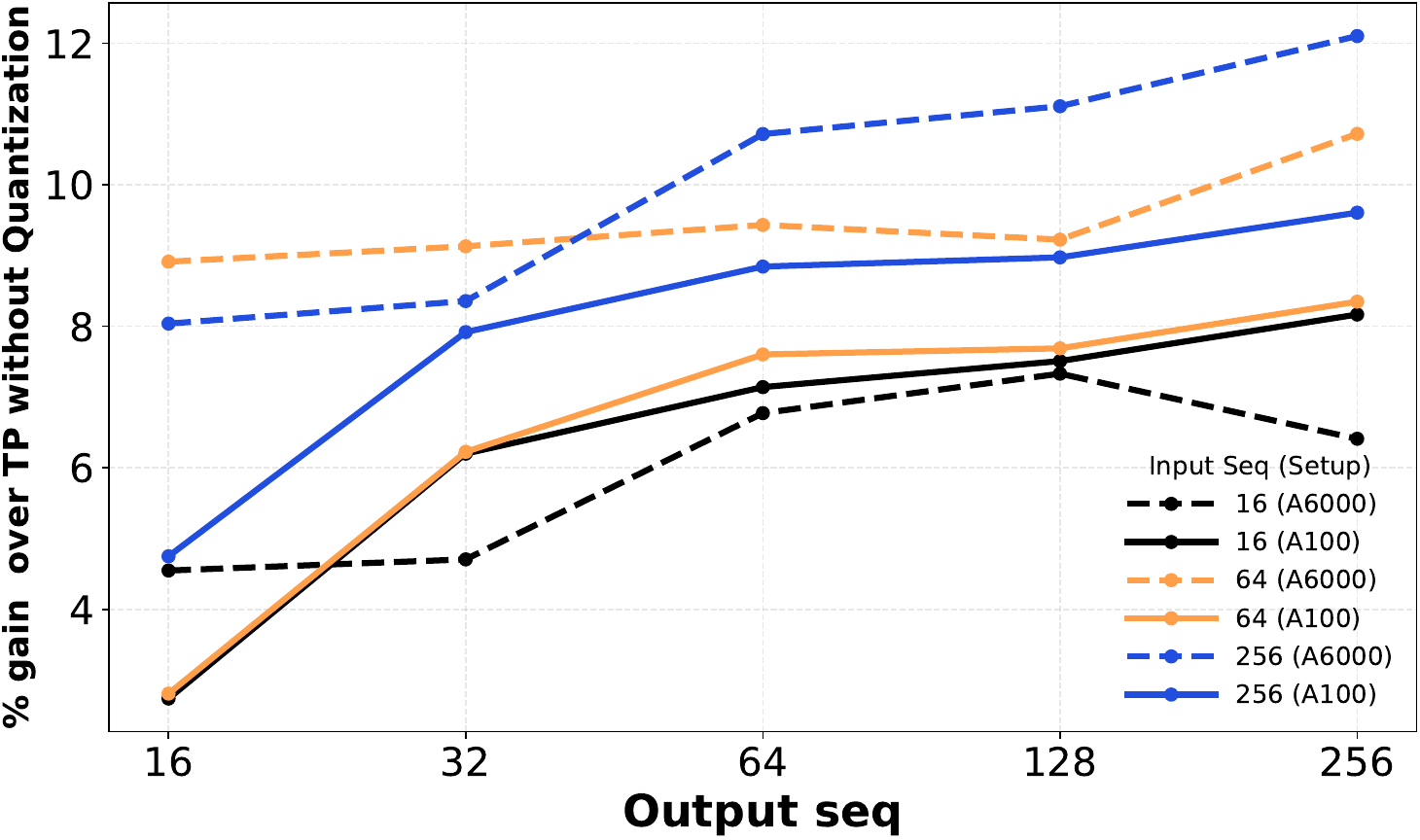}
  \caption{Throughput gains afforded by quantizing AllReduce in our TP inference design for Mamba as a function of output sequence length.}
  \label{fig:mamba_allreduce_quant}
  \vspace{-20pt}
\end{figure}

%
Figure~\ref{fig:mamba_allreduce_quant} shows the performance improvement afforded by quantizing AllReduce in our TP design for the Mamba model.
Specifically, we plot the throughput gain achieved by our TP design with quantization over TP without quantization.
We see that, across all input/output sequence lengths, quantization improves throughput by around 3--12\% on A6000 and by around 3--10\% on A100, with gains generally increasing with sequence length (larger activations and more collective traffic), peaking at 12.1\% for input and output size of 256 on A6000 and 9.6\% at the same point on A100. 
On closer inspection, we find that the (latency) improvement is most pronounced for TTFT where collective synchronization lies on the critical path.
On the PCIe-connected A6000, the gains (dotted lines) remain more clearly differentiated across input lengths because AllReduce constitutes a larger fraction of end-to-end latency, whereas on the NVLINK-connected A100 the gains (solid lines) are more similar as faster collectives reduce the available headroom. This suggests quantized AllReduce becomes increasingly valuable as the communication fabric slows (e.g., PCIe, InfiniBand, multi-node deployments) since TP is often communication-bound and reducing collective payload size directly alleviates the dominant bottleneck.
With AllReduce quantization, we observe 12--18\% throughput gains for Falcon-Mamba and Zamba, with an average top-1 accuracy drop of 2.5\% and top-5 drop of 1.5\%.

\vspace{-10pt}
\subsection{Ablation study to examine the benefits of our TP design}
\label{ss:ablation}
\begin{figure}[t]
    \centering
    \includegraphics[width=0.9\linewidth]{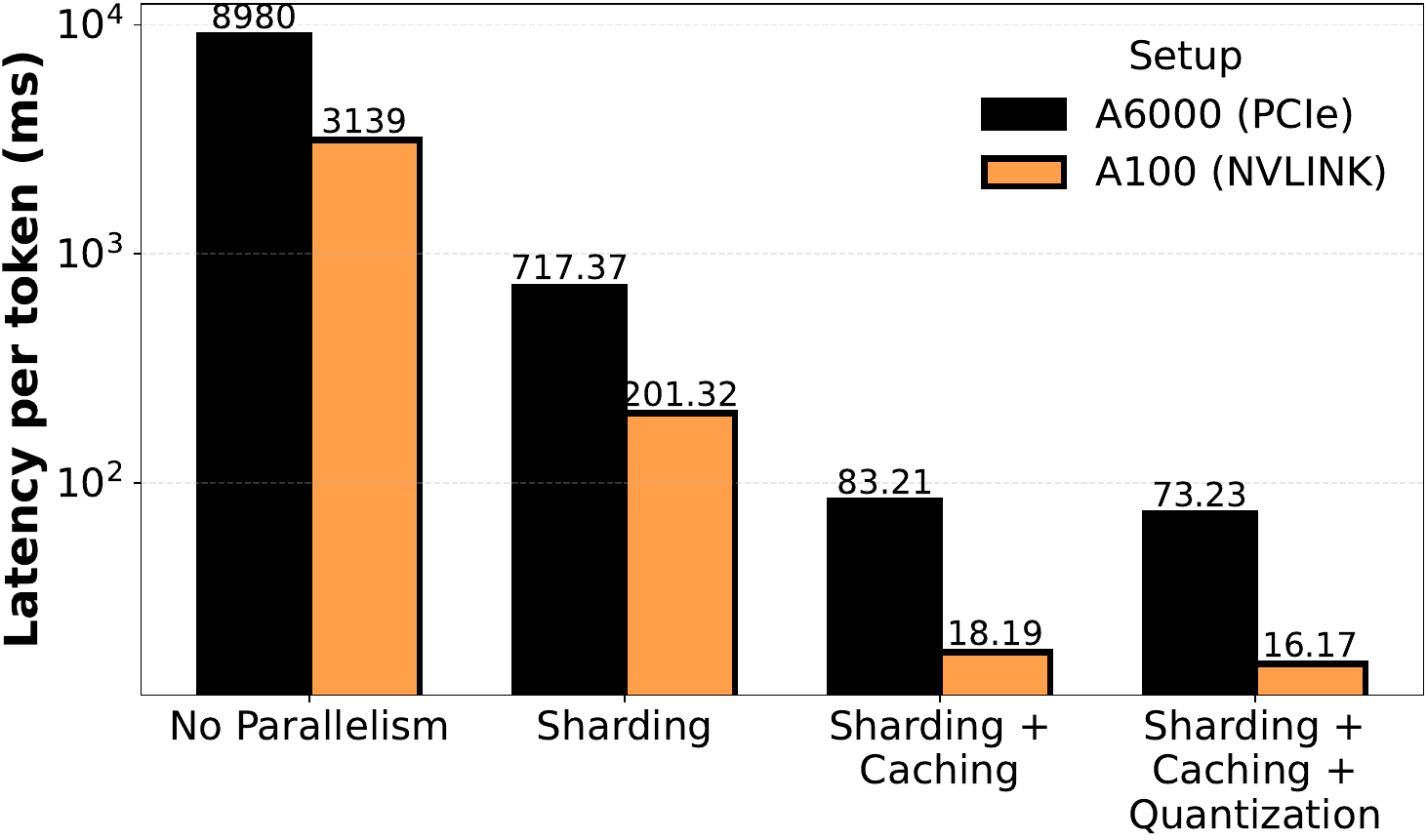}
    \caption{Ablation results for Mamba (256 input/output length).} 
    \label{fig:ablation_latency}
    \vspace{-0.6cm}
\end{figure}

To attribute the throughput gains afforded by our TP design (described in Section~\ref{sec:Design}), we now perform an ablation study.
Figure~\ref{fig:ablation_latency} shows the latency per token (on log scale) for the Mamba model at 256-token input and output lengths under our TP design; the conclusions are qualitatively similar at other lengths and for other models. 
The x-axis lists the incremental configurations we designed: only Sharding, Sharding with Caching, and Sharding, Caching, and Quantization. Note that our results in Section~\ref{ss:results} employed the Sharding + Caching design and those in Section~\ref{ss:quantization} employed the Sharding + Caching + Quantization design.

We see that both Sharding and Caching provide significant benefits.
On the A6000 GPUs, our sharding scheme reduces latency from about 9s to 717.4\,ms, a 12.5$\times$ improvement; introducing our SSM cache further reduces latency to 83.2\,ms, representing an $\sim$8.6$\times$ improvement over just sharding by avoiding redundant prompt re-processing during decode. Adding quantized AllReduce further decreases latency to 73.2\,ms ($\sim$12\% additional reduction), confirming that bandwidth-optimized synchronization yields measurable benefits even under PCIe interconnects. On the A100 (NVLINK) setup, the same trend holds: sharding alone reduces latency from about 3.1s to 201.3\,ms ($\sim$15.6$\times$); caching then reduces latency from 201.3\,ms to 18.2\,ms ($\sim$11$\times$), and quantized AllReduce brings it down further to 16.2\,ms ($\sim$11\% reduction). 
These results reinforce that caching primarily addresses redundant prefill compute, sharding preserves mixer locality and avoids extra synchronization, while quantized AllReduce mitigates communication bottlenecks on the critical path, with stronger relative gains on slower interconnects.


\section{Conclusions}
\label{s:conc}
This paper presents the first tensor-parallel design for SSM inference and its implementation for four popular SSMs (Mamba, Mamba2, Falcon-Mamba, and Zamba).
Our design overcomes the systems challenges of parallelizing SSMs by intelligent channel-wise model splitting to reduce synchronization, introducing an SSM cache to reuse recurrent state, and by implementing AllReduce quantization to further alleviate inter-GPU communication overhead. 
Our experimental results on two GPU platforms shows that our tensor-parallel SSM inference significantly improves throughput ($\sim$1.4--3.9$\times$) compared to the default, single-GPU inference. 
Importantly, our design allows us to handle larger SSM configurations, including $\sim$2--4.0$\times$ larger batch sizes and output sequence lengths, due to efficient memory usage under our tensor parallel implementation.

\bibliographystyle{IEEEtran}
\bibliography{references}


\end{document}